\documentclass[%
 reprint, amsmath,amssymb,aps,pra]{revtex4-2}
\usepackage{bm} 
\usepackage{graphicx}
\usepackage{subcaption}
\usepackage{dcolumn}
\usepackage{bm}
\usepackage{braket}
\usepackage{hyperref}
\usepackage[mathlines]{lineno}

\begin{document}

\preprint{APS/123-QED}

\title{Simple Vector Magnetometer Based on Ground State Hanle Effect}

\author{Nayan Sharma}
\email{nlop2022@gmail.com}
\affiliation{Department of Physics, Sikkim University, 6th Mile Samdur, East Sikkim, India -737102}
\author{Ranjit Kumar Singh}
\affiliation{Department of Physics, Sikkim University, 6th Mile Samdur, East Sikkim, India -737102}
\author{Ajay Tripathi}
\affiliation{Department of Physics, Sikkim University, 6th Mile Samdur, East Sikkim, India -737102}

\begin{abstract}
We present a method for determining the azimuthal phase (angle) of a magnetic field by exploiting phase matching of laser beams in a ground-state Hanle effect (GHSE) configuration. This approach is based on the symmetry of the system's Hamiltonian and the existence of a phase-independent frame, allowing for direct determination of the field orientation. As a proof of concept, we performed preliminary experiments using the F=1$\rightarrow$F=0 transition of the D$_2$ line in $^{87}$Rb, with three laser beams to demonstrate the phase (azimuthal) dependence of the observed Hanle resonance signals. While our current setup does not include active phase control, the key features predicted by our method were observed, validating its conceptual foundation. Additionally, we measured two components of the stray magnetic field in our laboratory as an illustration. This method leverages the Hanle effect’s inherent sensitivity to both the magnitude and orientation of magnetic fields, as well as the underlying symmetry properties of the atomic system, and offers a pathway for precise, calibration-free determination of magnetic field orientation.

\end{abstract}

\maketitle

\section{Introduction}
Accurate measurement of magnetic fields is essential in atomic physics, quantum optics, and many related fields.  Recent advances in atomic magnetometry have made it possible to achieve highly sensitive and versatile measurements of magnetic fields using thermal alkali vapor cells. The ground-state Hanle effect (GHSE) ~\cite{breschi2012ground} is a widely used technique for detecting arbitrary magnetic fields, making it a valuable tool for magnetometry.\\
In typical experiments with GHSE using $\pi$ polarized light, it is observed that the transmission (or absorption) resonance splits when a transverse magnetic field is present ~\cite{margalit2013degenerate} .This splitting varies linearly with the strength of the transverse magnetic field. However, this observed splitting is also sensitive to factors such as the collisional decay of ground-state spin polarization and the range over which the magnetic field is scanned. To reliably measure very small magnetic fields (less than a milligauss), atomic vapor cells are often used with buffer gases and anti-relaxation coatings such as octadecyltrichlorosilane (OTS). These features help to reduce spin relaxation and enhance the sensitivity and resolution of the magnetometer in low-field regimes. Full vector reconstruction typically employs intensity modulation ~\cite{huang2015three}, external bias fields ~\cite{schonau2025optically}, Stark shift detection ~\cite{patton2014all}, among other methods.\\
In this work, we present a method to the measure the azimuthal angle of an arbitrary magnetic field vector using a $\pi$-polarized probe beam. To achieve this, we need to introduce two additional laser beams of circular polarizations, each with independent phase control. Theoretically we show that by leveraging symmetry arguments how tuning the relative phases of these beams allows for direct determination of the magnetic field’s azimuthal angle. As a proof of concept,  we also present an experiement to illustrate the main features of the observed asymmetry when multiple beams are used without phase control.

\section{Theoretical Framework}
We consider a specific configuration in which three coherent light fields, each having the same frequency, propagate along the x-axis. The electric field components of the light fields are given by,
\begin{eqnarray}
	\bm{E}_p = E^0_{p}\text{cos}(\omega t + \phi_p)\hat{z},~~~ \bm{E}_{\pm}=E^0_{\pm}\text{cos}(\omega t + \phi_{\pm})\hat{e}_{\pm} 
\end{eqnarray}

where $E_p$ is the probe field and $E_{\pm}$ are the control fields. The phases $\phi_p$ and $\phi_{\pm}$ correspond to the probe and control fields, respectively. The probe field $\bm{E}_p $ is linearly polarized along the $z$-axis, while the control fields $\bm{E}_{\pm}$ are circularly polarized, with their polarization basis vectors defined as $\sqrt{2}\hat{e}{\pm} = \hat{y} \pm i \hat{z}$. The total magnetic field is the sum of a static three-dimensional vector and a time-dependent scanning field along the z-axis:
\begin{eqnarray}
	\bm{B} &=& B_x ~\hat{x} +  B_y ~ \hat{y} + [B_z + B_s] ~ \hat{z}
\end{eqnarray}
where $B_x$, $B_y$, and $B_z$ are the static components , and $B_s$ is the scanning field in the z-direction.
These fields interact with a four-level atomic system configured in a closed-loop tripod scheme, where the total angular momentum of the ground state is $F = 1$ and that of the excited state is $F' = 0$.\\

The interaction Hamiltonian ($\hbar=1$) under the rotating wave approximation (RWA), expressed in a time-independent basis for resonant fields, is given by

\begin{eqnarray}
	\nonumber
	H_I =\frac{e^{i \pi}}{2}\left(\Omega_+ e^{i \phi_+} \ket{1}\bra{4} +\Omega_1 e^{i \phi_p} \ket{2}\bra{4} +\Omega_- e^{i \phi_-}\ket{3}\bra{4}\right)\\
	\nonumber
	+ h.c.
\end{eqnarray}
where, $\Omega_{\pm} = c_{\pm} E^0_{\pm}d_{12}$ and $\Omega_p=c_{p}E^0_1d_{12}$ are the Rabi-frequency coupling the ground and excited state expressed in terms of reduced dipole matrix element $d_{12}$ and transition strengths $c_p$, $c_{\pm}$.
\begin{figure}[t]
	\centering
	\includegraphics[width=0.5\textwidth]{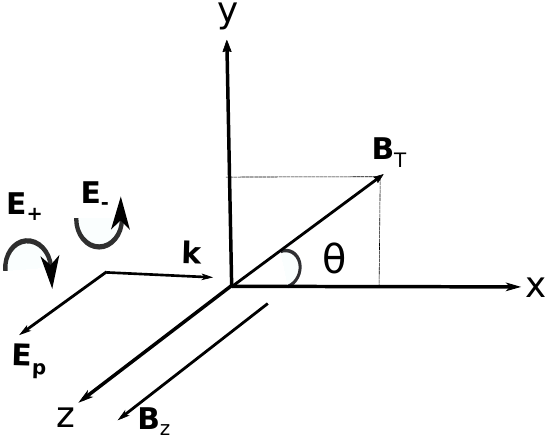}
	\caption{Schematic showing the geometry of the problem.}
	\label{coordinate}
\end{figure}
The Hamiltonian due to interaction with magnetic field is given by,

\begin{eqnarray}
	H_B = \Delta(\ket{1}\bra{1}-\ket{3}\bra{3}) \\
	\nonumber
	+ \left[\frac{\Omega_T}{\sqrt{2}} e^{-i\theta}\left(\ket{1}\bra{2}+\ket{2}\bra{3}\right)+h.c.\right]
\end{eqnarray}

  where $\Delta = g_{F} \mu_{B} (B_z + B_s)$ is the Larmor precession frequency associated with the total magnetic field along the $z$-direction, including both the static ($B_z$) and scanning ($B_s$) components. Here, $g_F$ is the Landé g-factor and $\mu_B$ is the Bohr magneton.  $\Omega_T = g_{F} \mu_{B} B_{T} $ is the Larmor precession frequency due to the transverse field $B_{T}=\sqrt{B^2_x + B^2_y}$. This transverse field couples ground states with $\Delta m = \pm 1$. The angle of the transverse field in the plane is defined by  $\text{tan}\theta =\frac{B_y}{B_x}$.The z-axis is chosen as the quantization axis for this problem, which determines the selection rules and the form of the coupling.\\
 
To study the optical response of the light fields, we adopt the simplest possible decay mechanism, represented by a diagonal matrix. 

\begin{eqnarray}
	\Lambda =	\gamma \sum_{i=1}^{4} \ket{i}\bra{i} + \Gamma \ket{4}\bra{4}
\end{eqnarray}

Here, $\gamma = \tau^{-1}$ represents the transit-time decay rate, where $\tau$ is the average time an atom spends within the interaction region, given by the beam diameter divided by the average atomic velocity. $\Gamma$ is the spontaneous emission rate associated with the decay of the excited state. A repopulation matrix is also introduced to ensure conservation of population and to study the steady-state behavior of the system

\begin{eqnarray}
	R =	\frac{\gamma}{3} \sum_{i=1}^{3} \ket{i}\bra{i} + \frac{\Gamma}{4} (\ket{1}\bra{1} +\ket{3}\bra{3}+2\ket{2}\bra{2} )
\end{eqnarray}
It is assumed that repopulation due to transit-time decay is primarily governed by the thermal velocity distribution of the atoms. For spontaneous decay processes, the repopulation of ground states is determined by the branching ratios of the transitions.\\
The optical Bloch equation is given by
 \begin{eqnarray}
 	\frac{\partial\rho(t)}{\partial t} = -i [H,\rho] - \frac{1}{2} \{\Lambda,\rho\} + R
 \end{eqnarray}
 and is analyzed under steady-state conditions. While approximate analytical solutions can be obtained in certain parameter regimes, in this work we perform a numerical study to achieve greater accuracy and convergence across a broad range of parameters.\\

\begin{figure}[t]
	\centering
	\includegraphics[width=0.5\textwidth]{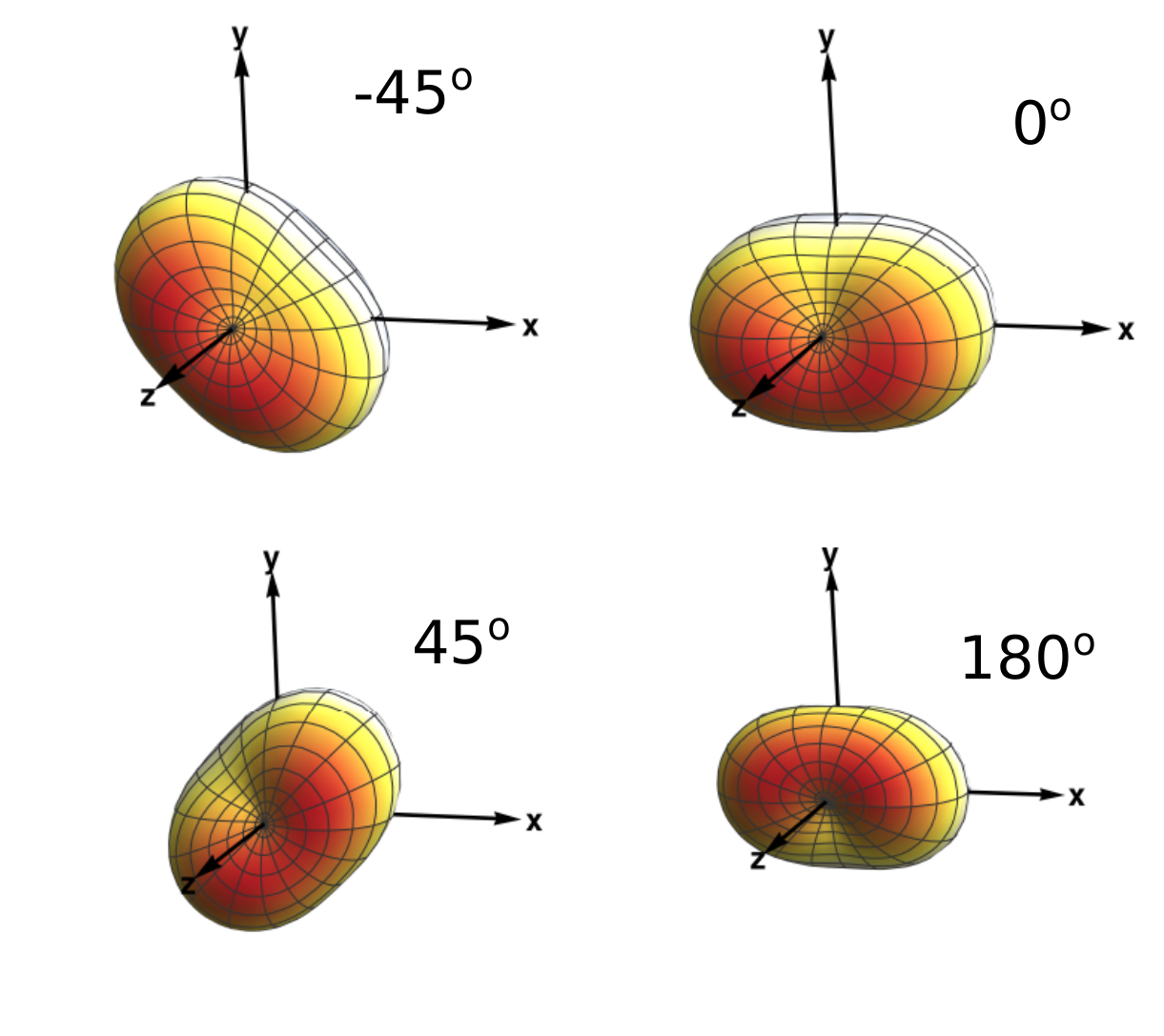}
	\caption{Angular Momentum Probabilty Surface (AMPS) for the single beam case.}
	\label{amps}
\end{figure}

\subsection{Single Beam ($\Omega_+=0=\Omega_-$): Phase Independent}
\begin{figure}[t]
	\centering
	\includegraphics[width=0.5\textwidth]{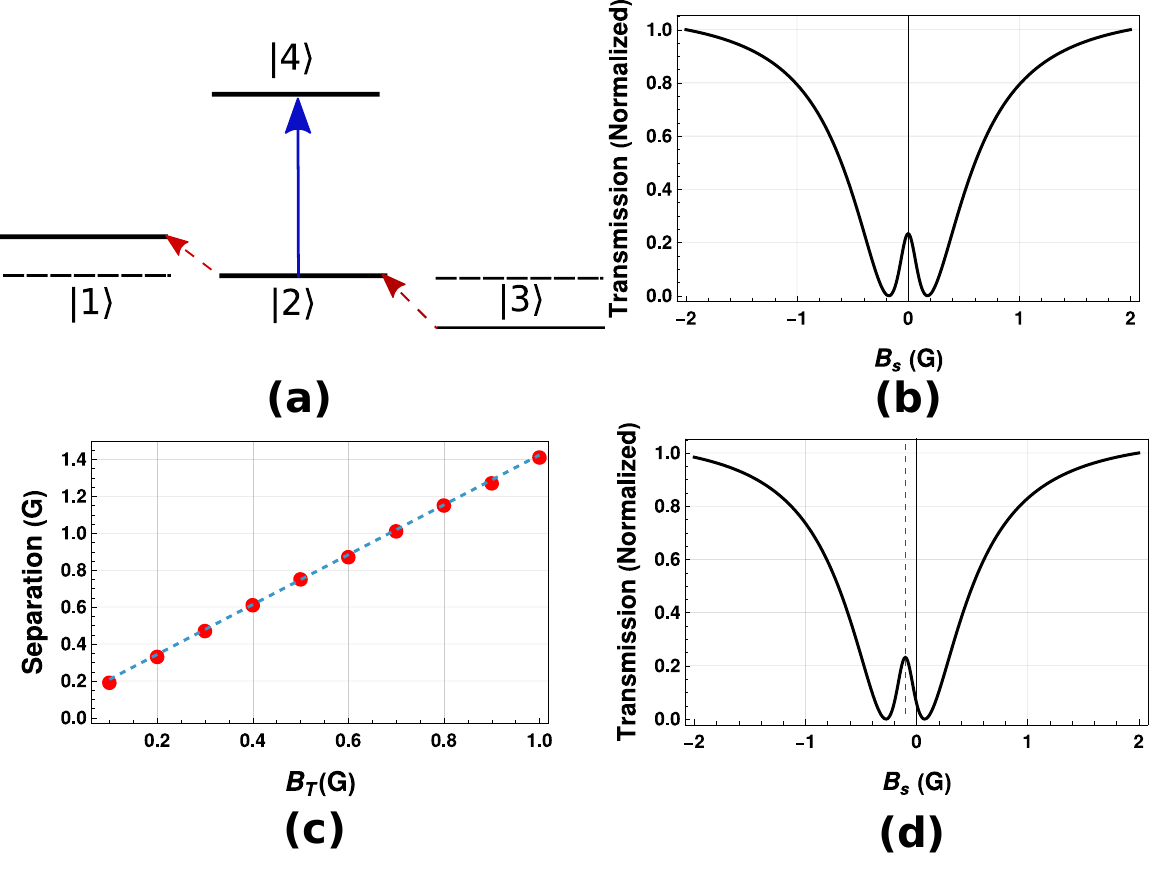}
	\caption{(a) Engery level diagram for single beam with a $\pi$ polarized probe. (b). Transmission v/s Scanning magnetic field. (c) Variation of the speration between two dips v/s the magnitude of the transverse magnetic field. (d)Transmission spectra showing the shift in the position of the central resonance which is directly proportional to the static $B_z$ component of the magnetic field.}
	\label{singlebeam}
\end{figure}

The energy level diagram for this single-beam configuration is depicted in Fig.~\ref{singlebeam}(a). In this setup, the linearly polarized light directly couples the two magnetic field-insensitive states, $\ket{2}$ and $\ket{4}$. Additionally, because the quantization axis is chosen along the $z$-direction, the ground states experience additional coupling mediated by the transverse magnetic field, which introduces a phase factor $\theta$ into the dynamics.\\
If we write the total Hamiltonian as $H = H_I + H_B$ for this case, it is always possible to construct a unitary transformation,
\begin{eqnarray}
	U = \mathrm{diag}(e^{i \theta}, 1, e^{- i \theta}, e^{i \phi_p})
\end{eqnarray}
that transforms the system into a phase-independent basis. This transformation does not alter the underlying physics or dynamics of the system, but simply removes explicit phase dependence from the Hamiltonian representation. This implies that, in the single beam ($\pi$ polarized) configuration, it is not possible to extract information about the angle $\theta$, as any dependence on $\theta$ can be removed by a suitable unitary transformation. As a result, the system's observables are insensitive to $\theta$ in this case. This $\theta$-independence can also be understood by examining the angular momentum probability surfaces (AMPS) for the ground states, as illustrated in Fig.~\ref{amps}. A change in $\theta$ corresponds to a rotation of the AMPS about the $z$-axis. This reflects the well-known axial symmetry of the system: physical properties remain invariant under rotations around the quantization axis. Consequently, any coherences or populations probed along the $\hat{z}$ axis remain unaffected by $\theta$, reinforcing the phase insensitivity in this configuration.\\
Although the system exhibits axial symmetry, the transmission of the probe beam depends on the magnitude of the transverse magnetic field $B_T$. The presence of $B_T$ gives rise to a Hanle resonance, which is characterized by an increase in transmission (gain) at the line center, flanked by two minima. This central gain feature, appearing between two dips in the transmission spectrum as a function of magnetic field, is a hallmark of the Hanle effect in such systems. This effect is illustrated in Fig.~\ref{singlebeam}(b). For this plot, the static magnetic field component $B_z$ was set to zero to study of the effect of transverse field. As reported previously, we find that the separation between the two dips in the the transmission spectrum increases increases linearly with the magnitude of the transverse magnetic $B_T$ (see Fig.~\ref{singlebeam}(c)). Fig.~\ref{singlebeam}(d) shows the effect of $B_z$, which shifts the resonance by an amount proportional to $B_z$. Therefore, in principle two components of a unknown magnetic field can be found using this single beam setup.\\

\subsection{Multi-Beam: Phase dependent}
\begin{figure}[t]
	\centering
	\includegraphics[width=0.5\textwidth]{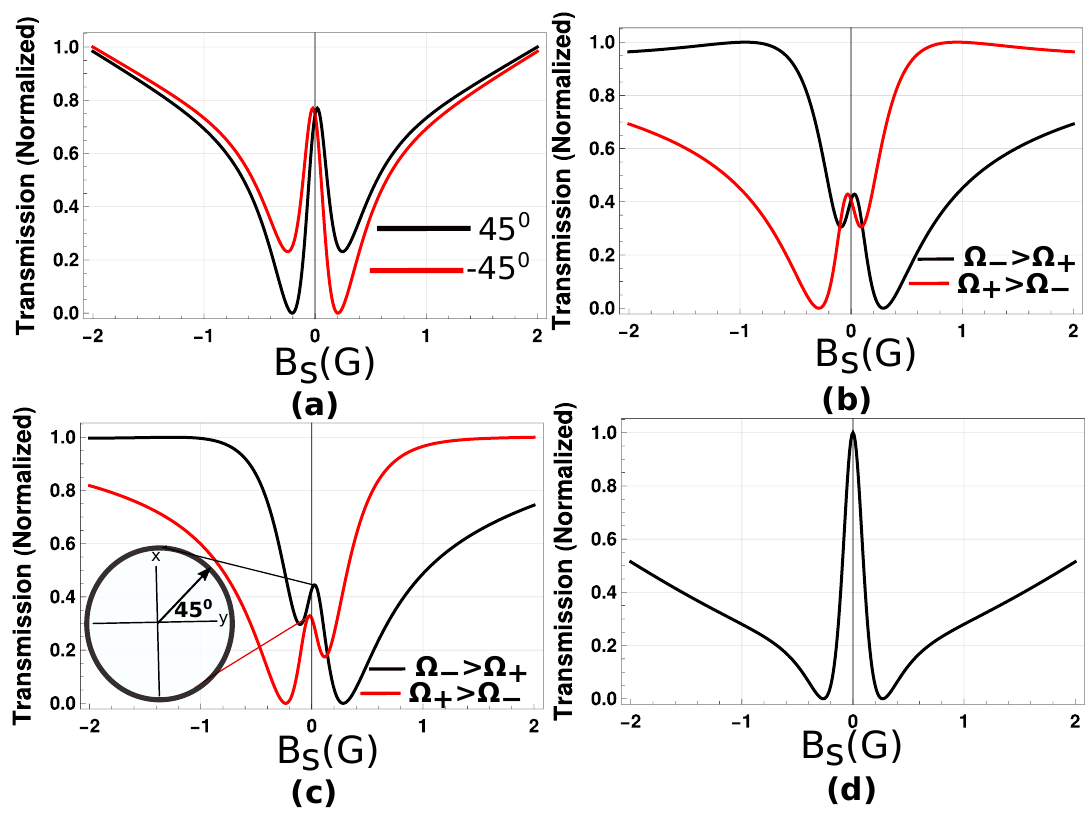}
	\caption{Probe transmission spectra for (a). two different values of $\theta$ for multi beam case, (b). $\theta=0$ and unequal powers of the $\sigma^{\pm}$ beams, (c). $\theta=0$ and unequal powers of the $\sigma^{\pm}$ beams, (d). $\theta=0$ and equal powers of the $\sigma^{\pm}$ beams. }
	\label{doublebeam}
\end{figure}

\begin{figure*}[htp]
	\centering
	\includegraphics[width=0.9\textwidth]{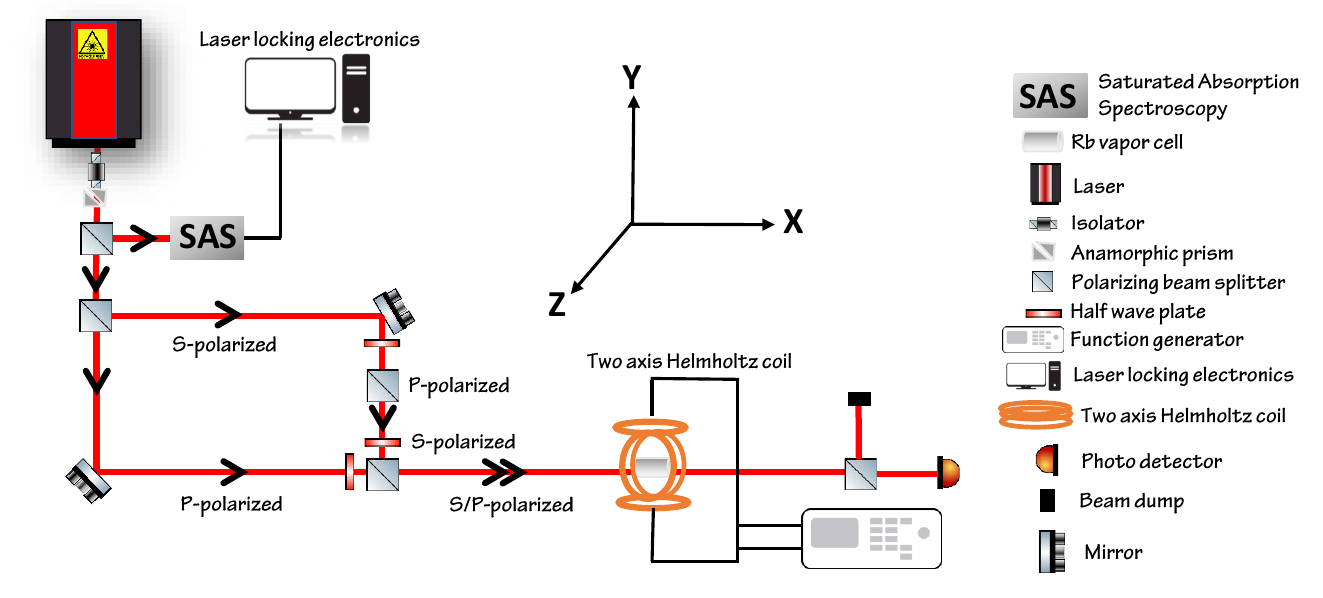}
	\caption{Schematic of the experimental setup used in the experiment.}
	\label{elevel}
\end{figure*}

For the case of three light beams, the symmetry of the Hamiltonian does not allow transformation to a phase-independent basis unless the following  phase matching condition is satisfied,
\begin{eqnarray}
	\phi_1 - \phi_+ = \theta\\
	\phi_1 - \phi_- = - \theta
\end{eqnarray}
This is also true if only two beams $E_{\pm}$ are used. Throughout this discussion we are interested to study the change in the optical response of the $\pi$ polarized beam $E_p$, the Rabi frequencies of the additional beams are always kept less than that of the probe ($\Omega_+, \Omega_- < \Omega_p$), so that these beams only slightly perturb the system without significantly altering the spectral features. This allows us to specifically study how the optical response of the central maximum and its surrounding minima changes under weak perturbation. Figure~\ref{doublebeam}(a) shows the probe transmission spectrum under phase-dependent conditions for $\Omega_+ = \Omega_-$. The two traces correspond to different $\theta$ values, revealing induced asymmetry around the central peak, where the two dips are no longer symmetric. The system also exhibits a different kind of asymmetry when the Rabi frequencies of the control beams are unequal, as shown in Fig.~\ref{doublebeam}(b). In this case, all phases are set to zero to isolate the effect of unequal beam powers. In experiments, this scenario can be realized by changing the polarization of linearly polarized light—typically by using a quarter-wave plate to divide the light into different circular polarization components.
Fig.~\ref{doublebeam}(d) shows the combined effect of unequal Rabi frequencies and varying $\theta$ in the phase-dependent system. We observe a shift in the amplitude of the central resonance as $\theta$ is introduced, highlighting the interplay between phase and power imbalance in shaping the spectral response.\\
In a phase-matched scenario, where the system can be transformed to a phase-independent basis, the response of the probe beam closely resembles that of the single-beam case in terms of symmetry. Although the presence of the control beams can modify the amplitude of the central resonance, the overall spectral lineshape remains symmetric about the central peak (see Fig.~\ref{doublebeam}(d)).\\
If we consider a case where the angle $\theta$ between $B_x$ and $B_y$ is fixed but unknown, the system generally exhibits asymmetry in its response. By tuning the relative phases of the beams—either optically or electronically—the system can be brought back to a symmetric state when the phase matching condition is achieved. This restoration of symmetry provides a direct method to determine the value of the unknown angle $\theta$ which can be the third component of the unknown magnetic field.\\

Below we present some preliminary experiments performed on the $D_2$ line of $^{87}$Rb vapor, where we observe the key features predicted by our theoretical framework. Although a full-scale three-dimensional vector reconstruction was not possible due to the lack of precise phase control in our current setup, we present  the measurement of two component of lab stray fields using single beam. We also report the observation of the asymmetry in case of multi beams and how that asymmertry vanishes if the stray field is compensated such that the angle $\theta$ is made zero. This work serves as a proof-of-principle demonstration; we have not addressed sensitivity or optimization, but rather focused on illustrating the feasibility of the approach.

\section{Experiment}\label{exp}

\subsection{Experimental setup}
The schematic of the experimental setup used in the experiments is shown in Fig. \ref{elevel}. A single external cavity diode laser (ECDL) operating at 780 nm is used as the source. The laser frequency is stabilized via saturated absorption spectroscopy (SAS) using a reference Rb vapor cell and Toptica Digilock 110 laser locking module.

The laser output is first passed through an isolator to minimize back reflections, and then through an anamorphic prism pair to circularize the beam profile. The beam is split into two orthogonally polarized components using a polarizing beam splitter (PBS). These components are directed through appropriate optical paths and recombined using another PBS after polarization adjustment using half-wave plates. The resulting beam thus contains both S- and P-polarized components, forming the probe and control beams. The beam diameter for both the beam is around 0.2 cm. For the experiments, the probe and control power is maintained at 5 $\mu$W and 1 mW, respectively. The propagation geometry is co-linear, with both probe and control beams overlapping spatially in the Rb cell.
The combined beam passes through a Rb vapor cell positioned at the center of a two-axis Helmholtz coil system, which enables the generation of a uniform magnetic field along controlled directions. The magnetic field is scanned by driving the coils with a function generator. After interaction with the vapor, the transmitted light is detected by a fast photodetector and measured on a digital oscilloscope. To calibrate the experimental data, the ramp signal generated by the function generator which is used to scan the magnetic field—is simultaneously recorded along with the transmission signal.

\subsection{Results and discussion}

\begin{figure}[htp]
	\centering
	\includegraphics[width=0.5\textwidth]{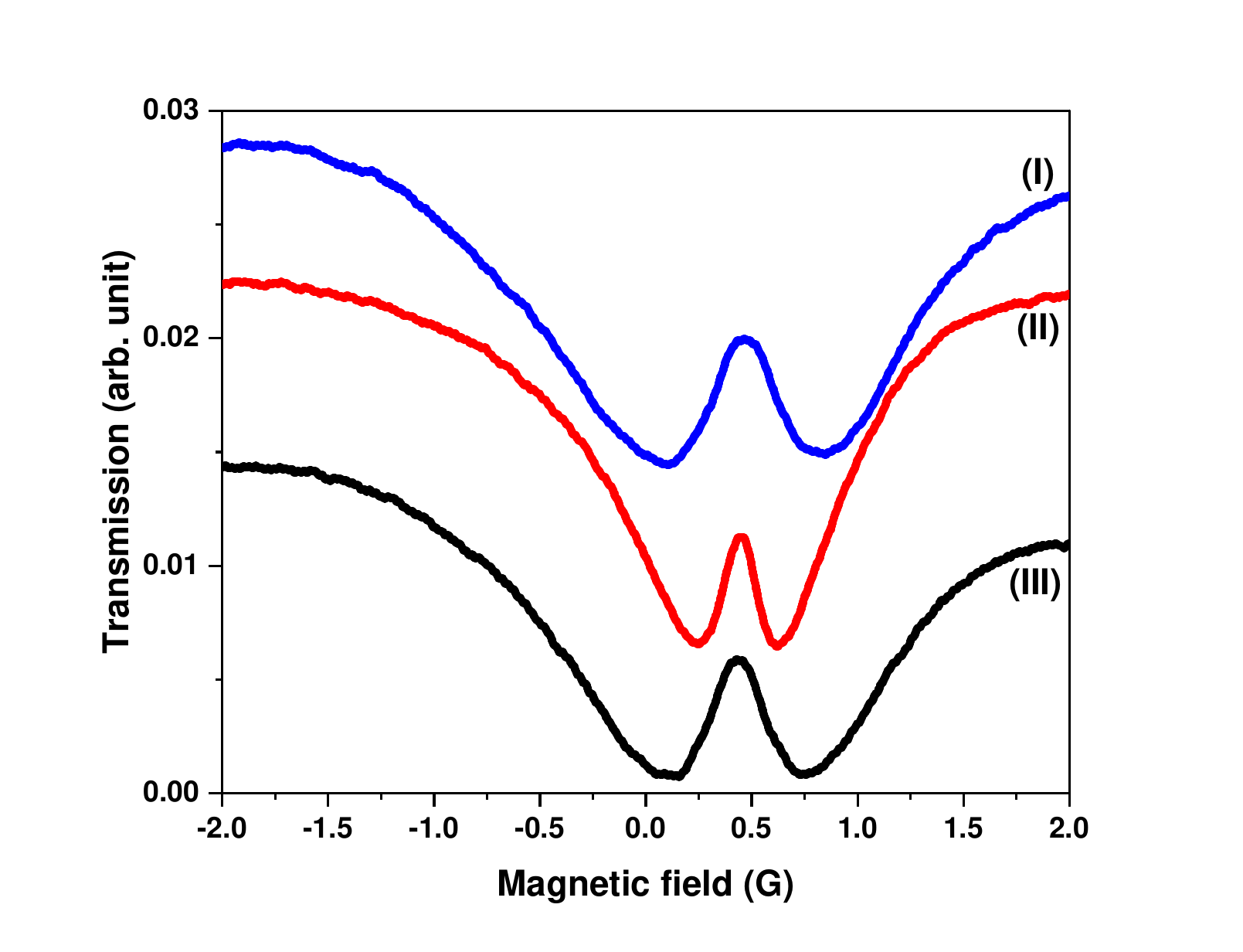}
	\caption{Probe Transmission v/s scanning magnetic field for single beam case. Probe power =100$\mu$W.}
	\label{exptsingle}
\end{figure}

Figure \ref{exptsingle} shows the Hanle resonances observed in the single-beam configuration wih lab stray fields. For this measurement, the probe power was kept fixed at 100$\mu$W.Two Helmholtz coils were used in the experiment, one coil was used to scan the magnetic field along the z-direction, while the other coil was used to generate a compensating magnetic field in the y-direction(By). Curve I corresponds to the case where the applied magnetic field By was zero. The observation of Hanle resonance at this condition predicted the presence of transverse component of the stray magnetic field. As the magnetic field along the second axis (By, in the positive y-direction) was increased from 0G, seperation between the resonance minima decreased and the minimum separation was found at approximately 0.2G(Curve II). Upon further increasing By, the separation between the minima increased again(Curve III).\\

\begin{figure}[htp]
	\centering
	\includegraphics[width=0.4\textwidth]{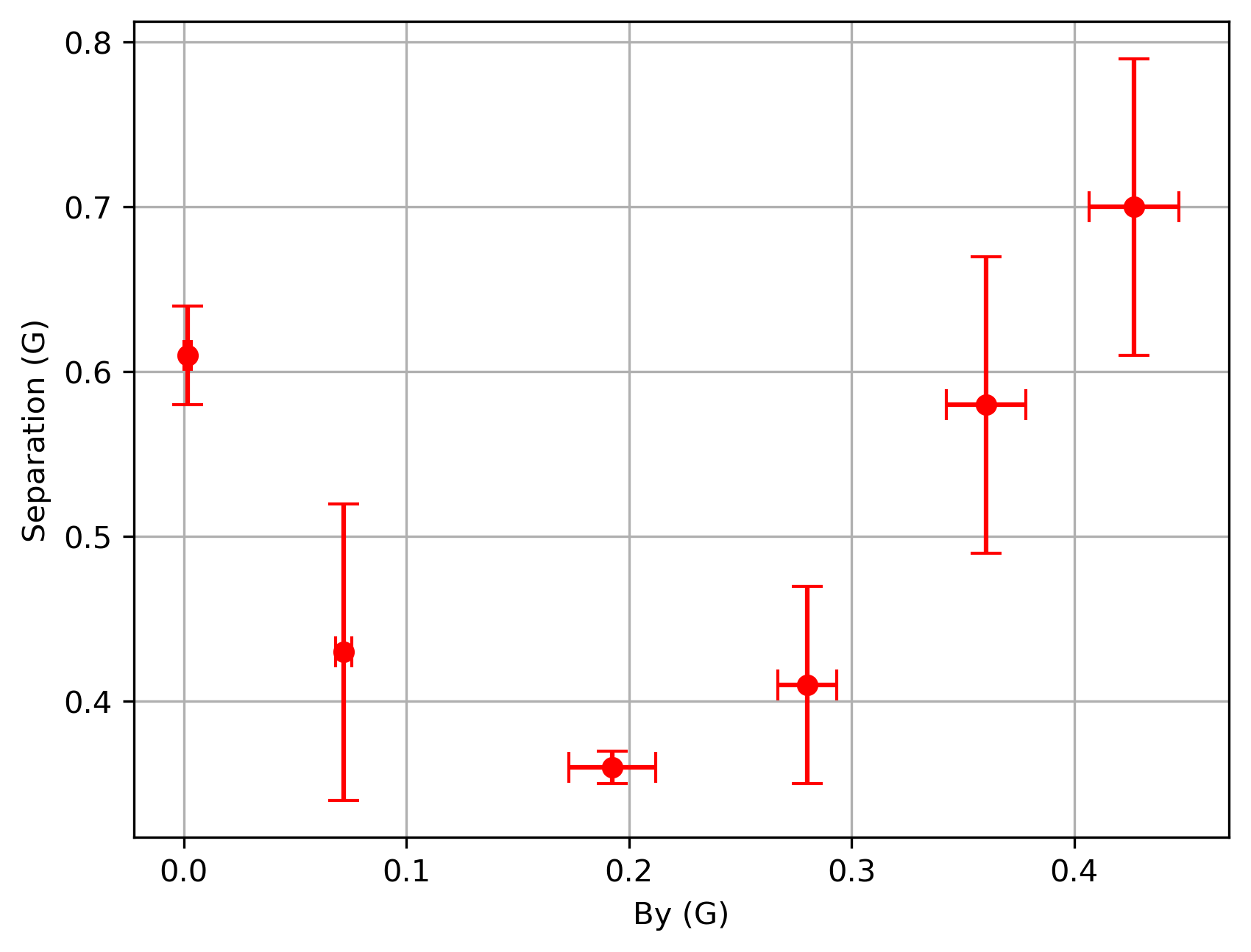}
	\caption{Separation v/s the compensating magnetic field.}
	\label{expterror}
\end{figure}

This trend was utilized to determine the y-component of the stray magnetic field present in our laboratory setup. The value of this y-component was found to be 0.192$\pm$0.019 G, as determined by identifying the value of By at which the minimum separation between the resonance minima was observed(see Fig.\ref{expterror}). The error bars along the y-axis reflect the uncertainty in measuring the separation, which was primarily due to noise and the limited number of measurements (only two data points were used in this case). The uncertainties along the x-axis were estimated by propagating the errors associated with voltage/current measurements (as measured by the digital multimeter) and the calibration uncertainty of the Helmholtz coil. Despite these limitations, a clear and consistent trend is evident, indicating the robustness of the compensation method.\\
For all the measurements the central maximum of the resonance curve was observed to be shifted from zero towards the positive side. As discussed previously, this shift indicates the presence of a z-component of the stray magnetic field, oriented along the -$\hat{z}$ direction in our case.
From our measurements, the magnitude of this shift was determined to be 0.427 $\pm$ 0.010 G, which corresponds to the value of the z-component of the stray field in our setup.
The direction and magnitude of the measured stray magnetic field components were found to be comparable to those of the Earth's magnetic field at our location. This observation indicates that, in the absence of magnetic shielding (such as $\mu$-metal), the current experimental setup is predominantly influenced by the Earth's magnetic field. In principle, the third component of the unknown magnetic field can be determined by employing a third Helmholtz coil along the corresponding axis. However, we propose a calibration-free method to fix this third component, which would be the azimuthal angle $\theta$.\\
\begin{figure}[htp]
	\centering
	\includegraphics[width=0.45\textwidth]{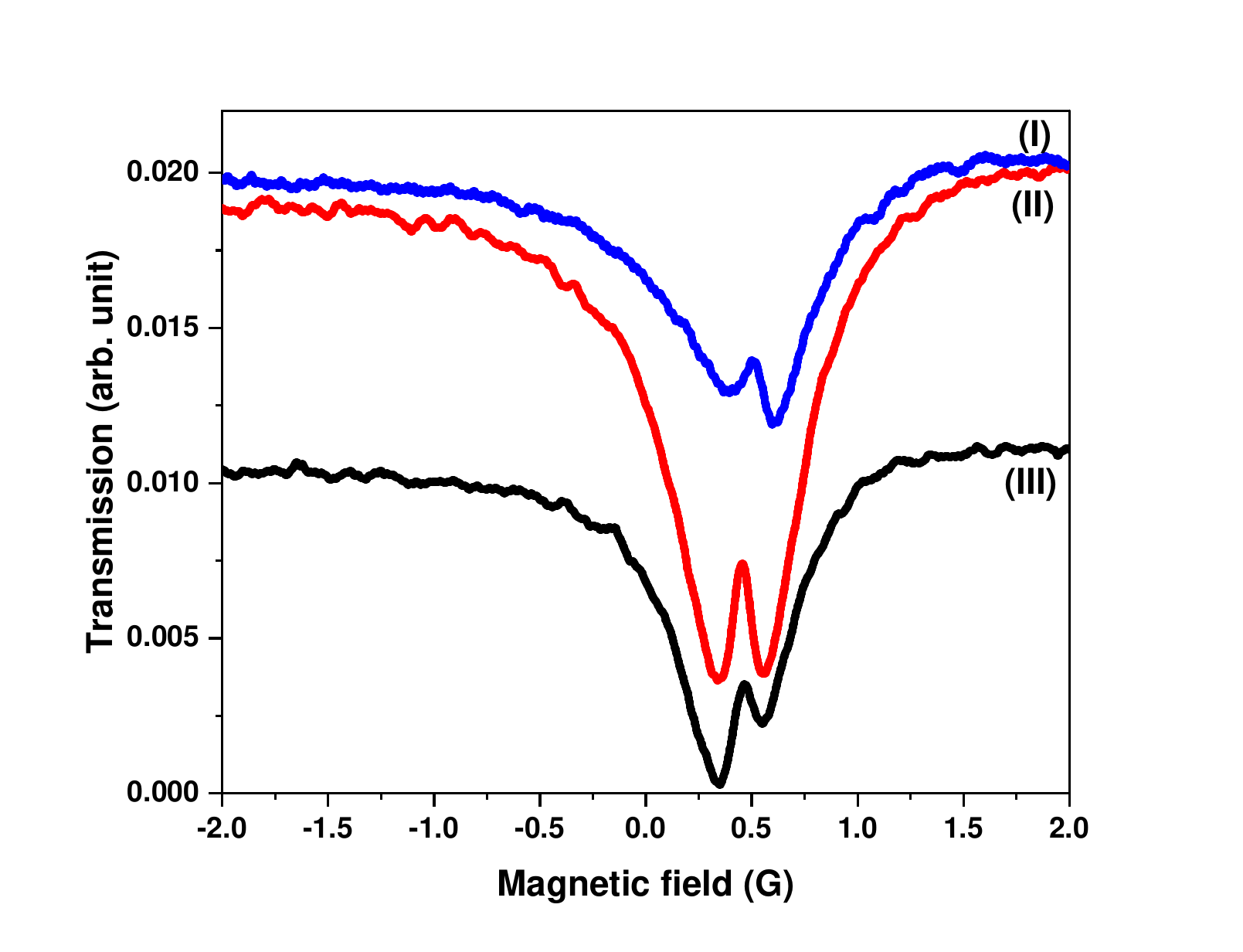}
	\caption{Probe Transmission v/s scanning magnetic field for multi-beam case. Probe power =100$\mu$W. Pump power = 50 $\mu$W}
	\label{exptdouble}
\end{figure}

 Figure \ref{exptdouble} presents the results obtained when control beams were introduced into the setup. The powers of the control beams were kept equal at 50$\mu$W.Curve I corresponds to the case when By=0, where a pronounced asymmetry is observed in the resonance profile. Applying a magnetic field By that compensates the y-component of the stray field, this asymmetry disappears, and the resonance lineshape closely resembles that of the single-beam configuration (see Curve II).At this compensation point, the the azimuthal angle is effectively set $\theta$=0, making the system $\theta$ independent. Further increasing By causes the asymmetry to reappear, but with opposite polarity as shown by Curve III. This indicates a transition of the azimuthal angle from one quadrant to another. While a detailed mapping of this asymmetry as a function of $\theta$ can be performed to measure the angle, such an approach requires precise calibration.
To overcome this limitation, we propose a calibration-free technique wherein the relative phase between the control beams is modulated to satisfy the phase-matching condition. Under this condition, the system becomes $\theta$ independent, and the asymmetry vanishes. Thus, the phase-matching condition provides a direct and calibration-free means to determine the azimuthal angle $\theta$.

\bibliography{reference}
\end{document}